\documentclass[sigconf, authorversion]{acmart}
\AtBeginDocument{%
  \providecommand\BibTeX{{%
    \normalfont B\kern-0.5em{\scshape i\kern-0.25em b}\kern-0.8em\TeX}}}

\usepackage{hyperref}

\usepackage{xcolor}
\usepackage{listings}

\newcommand{\change}[1]{{\textcolor{black}{#1}}}

\copyrightyear{2024}
\acmYear{2024}
\setcopyright{acmlicensed}\acmConference[IUI '24]{29th International Conference on Intelligent User Interfaces}{March 18--21, 2024}{Greenville, SC, USA}
\acmBooktitle{29th International Conference on Intelligent User Interfaces (IUI '24), March 18--21, 2024, Greenville, SC, USA}
\acmDOI{10.1145/3640543.3645200}
\acmISBN{979-8-4007-0508-3/24/03}

\begin{document}


\title[Why and When LLM-Based Assistants Can Go Wrong]{Why and When LLM-Based Assistants Can Go Wrong: Investigating the Effectiveness of Prompt-Based Interactions for Software Help-Seeking}

\renewcommand{\shortauthors}{}
\author{Anjali Khurana}
\email{anjali\_khurana@sfu.ca}
\orcid{0000-0002-2730-5512}
\affiliation{%
  \institution{Simon Fraser University}
  \state{BC}
  \country{Canada}
}
\author{Hari Subramonyam}
\email{harihars@stanford.edu}
\orcid{0000-0002-3450-0447}
\affiliation{%
  \institution{Stanford University}
  \city{}
  \state{}
  \country{USA}
}

\author{Parmit K Chilana}
\email{pchilana@cs.sfu.ca}
\orcid{0009-0007-0173-1752}
\affiliation{%
 \institution{Simon Fraser University}
  \state{BC}
  \country{Canada}
}
\renewcommand{\shortauthors}{Khurana et al.}

\begin{abstract}

Large Language Model (LLM) assistants, such as ChatGPT, have emerged as potential alternatives to search methods for helping users navigate complex, feature-rich software. LLMs use vast training data from domain-specific texts, software manuals, and code repositories to mimic human-like interactions, offering tailored assistance, including step-by-step instructions. In this work, we investigated LLM-generated software guidance through a within-subject experiment with 16 participants and follow-up interviews. We compared a baseline LLM assistant with an LLM optimized for particular software contexts, \textit{SoftAIBot}, which also offered guidelines for constructing appropriate prompts. We assessed task completion, perceived accuracy, relevance, and trust. Surprisingly, although SoftAIBot outperformed the baseline LLM, our results revealed no significant difference in LLM usage and user perceptions with or without prompt guidelines and the integration of domain context. Most users struggled to understand how the prompt's text related to the LLM's responses and often followed the LLM's suggestions verbatim, even if they were incorrect. This resulted in difficulties when using the LLM's advice for software tasks, leading to low task completion rates. Our detailed analysis also revealed that users remained unaware of inaccuracies in the LLM's responses, indicating a gap between their lack of software expertise and their ability to evaluate the LLM's assistance. With the growing push for designing domain-specific LLM assistants, we emphasize the importance of incorporating explainable, context-aware cues into LLMs to help users understand prompt-based interactions, identify biases, and maximize the utility of LLM assistants. 
\end{abstract}



\maketitle

\section{Introduction}

Learning to use feature-rich software applications for tasks such as advanced word processing, data analysis, image manipulation, and video editing can be challenging for end-users. Users currently turn to various software help resources to learn and seek help for such software tasks. For example, they usually begin by querying online search engines using keywords to locate specific resources, such as video and text-based tutorials, forums posts, and blogs and articles \cite{Kiani_help, Andrade_help, Novick_manual, Novick_tutorials}. However, online software-help seeking is a complex endeavour,  demanding precise queries to pinpoint the most pertinent information that can be directly applied within the application \cite{Kiani_help, Grossman_learnability}. 

The recent emergence of Generative AI and pre-trained Large Language Model (LLM)-based assistants like \textit{ChatGPT} \cite{AI_2022, brown_llm} offers a novel approach to support end-users' software help-seeking by leveraging these assistant's advanced natural language understanding \cite{Johnny}. For example, LLMs provide the potential for step-by-step instructions and explanations tailored to specific task needs, saving users time and effort searching through online resources. In the past two years, LLMs have demonstrated promising capabilities in assisting users in various domains, including tasks related to programming and software development  \cite{prompt_catalogue}, language generation \cite{AI_2022}, and question answering \cite{AI_2022}. However, the effectiveness of LLM-based assistance and how end-users employ LLMs to seek help for feature-rich applications remain important open questions. \change{Furthermore, there have been calls ~\cite{vera_liao} for a better understanding of the dynamics of user interaction across diverse AI usage scenarios, aiming to avoid false assumptions.} 

In this work, we investigate how software users make use of LLMs with \textit{prompt-based interactions} for seeking help for feature-rich applications. In particular, we investigate the effectiveness of recently emerging prompt guidelines \cite{Shieh_2023, Suhridpalsule_2023} offered by OpenAI, Microsoft, and others, which, when prepended to the user prompts, can potentially enhance LLM output to provide desired assistance \cite{Johnny}. Furthermore, we also consider the impact of integrating additional domain context, such as software documentation, into LLMs to improve the precision of the LLM output. For example, such techniques are currently being explored in in-application LLM assistants, such as Copilot in Microsoft 365 applications \cite{copilot}, Firefly in Adobe applications \cite{Adobe_2023}, etc.). For this investigation, we developed \textit{SoftAIBot}, our implementation of a state-of-the-art LLM assistant that integrates prompt guidelines and domain context, and compared it with a Baseline ChatGPT Plus LLM assistant. By analyzing users' interactions, behaviors, and challenges when using these state-of-the-art LLMs for software help-seeking, we aim to help end users harness the full potential of LLM-based assistants for feature-rich software \cite{MCCRACKEN_2023}. The research questions guiding this exploration were:

\begin{itemize}
    \item \textbf{RQ1:} How do end users make use of prompt-based interactions when finding software help ? 
    \item \textbf{RQ2: } To what extent can SoftAIBot generate accurate and relevant software help for end-users of feature-rich applications?
    \item \textbf{RQ3:} How do end users' mental models of LLMs influence their use of LLM-generated software help?  

\end{itemize}

In this paper, we report on the results from a controlled study and follow-up interviews that illustrate how end-users make use of prompt-based interactions when using LLM-based assistance in concert with tasks that involve visual interactions (e.g., Microsoft PowerPoint) and advanced data analysis/visualizations (e.g., Microsoft Excel). 
We ran a within-subject experiment with 16 participants from varied backgrounds without expertise in Machine Learning (ML) or Natural Processing Language (NLP). We hypothesized that our implementation of SoftAIBot with ChatGPT Plus (GPT-4) as the underlying LLM, state-of-the-art prompt guidelines, and integrated domain context will generate accurate and relevant assistance for users' prompts and help users finish their software tasks. However, our findings showed that even though SoftAIBot performed better than BaseLine in generating more accurate and relevant LLM output, users could not recognize these differences and struggled in mapping the LLM instructions to the software application, leading to poor task completion. The impact of the prompt text on the quality of the LLM output was not clear to users, and they rather followed LLMs' suggestions blindly, even when the output was inaccurate. Lacking an accurate mental model of LLMs, users tended to over-trust the LLM assistance without much contemplation.  

The main contributions of this research are in providing empirical insights that: (1) demonstrate how software users \change{employ new-generation LLM assistants to seek software help, both with and without prompt guidelines as well as integrated software context; }(2) illustrate the challenges that software users experience \change{with prompt-based interaction (e.g., crafting prompts, comprehending how prompts bias LLM output, mapping LLM-suggested steps to software, overtrusting output correctness)}; (3) identify gaps in users' mental models as they try to seek and apply assistance from LLMs to software tasks, \change{which affect both their use of LLMs and perception of software features, }regardless of implicit enhancements in the underlying model or explicit prompt guidelines \change{(SoftAIBot)}. We discuss the implications of these findings for designing and implementing LLM help tailored to feature-rich applications, the need to incorporate transparent and responsible LLM assistants, and ways to bridge the disparity between mental models and LLM interfaces, ultimately helping end-users form accurate mental models of LLMs.


\section{Related Work}
\label{related work}
This research drew upon insights from prior work on how users learn and seek help for feature-rich software, the emergence of LLMs for task-based assistance, and the use of prompt-based interactions.

\subsection{Software Help-seeking evolution}

HCI research has a rich history of investigating the challenges that users experience when learning and seeking help for complex feature-rich applications \cite{Grossman_learnability, Novick_tutorials, Kiani_help, Chilana_lemonaid}. 
Help-seeking resources and approaches have evolved over the years: from formal documentation and manuals \cite{Rettig, Novick_manual} to the use of videos \cite{Kim_videos, Lafreniere_Tutorials}, interactive tutorials \cite{Novick_tutorials}, Google Search, Q\&A or FAQ sites, blogs, dedicated forums \cite{Kiani_help} and even contextual help systems embedded within applications \cite{Brandt_joel, Chilana_lemonaid, Grossman_ToolClips, Hartmann, Lafreniere_incontext}. 

However, studies have shown that although users have increased access to help information, they often get lost in search results and forum posts and still face difficulty in recognizing effective and relevant \cite{Kiani_help, Novick} resources for completing their software tasks. Users also face numerous issues with articulating search queries using precise keywords \cite{Furnas, Kiani_help, Grossman_learnability}, often termed as \textit{vocabulary problem} \cite{Furnas}. Another related issue that users face is using the located help in coordination with the application and going back and forth between the two to accomplish their software tasks \cite{Kiani_help}. Past studies have shown that users tend to find step-by-step guidance within the context of feature-rich applications useful and trustworthy \cite{Khurana, Kiani_help, Chilana_lemonaid}. There have been constant innovations in devising help resources in the form of in-context help and video tutorials \cite{Lafreniere_Tutorials, Novick_tutorials, Grossman_ToolClips}. However, recent developments in Generative AI have opened a new outlet of help-seeking for users through LLM-based assistance, which is much more forgiving in letting users describe their queries using natural language \cite{Advait, Johnny}. These assistants tend to provide more specific and in-context assistance \cite{Johnny} to users, unlike traditional resources that may be scattered and require precise queries for retrieval. While there is a rich history of work supporting software learnability and easing help-seeking processes \cite{Chilana_lemonaid, Fourney, Grossman_ToolClips, Lafreniere_Tutorials}, it is unclear whether users’ learning approaches and strategies apply to LLM-based assistants. We extend this prior work by examining how novices employ LLM assistants for software help-seeking and the types of challenges they experience.

\subsection{LLM use for Task-Based Assistance} 

The emergence of powerful LLM assistants, such as ChatGPT, has significantly impacted task-based assistance for a range of domains, including programming, text generation, and text summarization \cite{tian2023chatgpt}. 
Recent studies have attempted to understand the use of LLM assistants for programming and software development-related help-seeking \cite{Barke, Xu}. For example, Xu, Vasilescu, \& Neubig (2022) \cite{Xu} investigated the use of LLMs for programming-related tasks and found that users struggle to generate assistance especially for complex queries as users struggled in formulating specific code-related input queries. Another study \cite{Vaithilingam} highlighted that users mostly rely on trial and error while debugging their code using LLM assistance and often do not feel confident about applying the output. 

 Recent interest in LLMs has inspired initiatives \cite{copilot, Adobe_2023} to utilize their capabilities for assisting users with software tasks as well \cite{copilot, Adobe_2023}. For example, some experimental work is being explored by integrating LLMs directly into feature-rich applications, such as Copilot in Microsoft 365 \cite{copilot} and Firefly in Adobe \cite{Adobe_2023}. This process has shown that developers can face new challenges in ensuring accurate and effective use of this new avenue of conversational UX experience \cite{vera_liao, Johnny}. As many of these interfaces are still at a nascent stage, it is unclear how this current practice (i.e., integrating the context of these feature-rich applications) can help novice end-users in seeking accurate and relevant assistance from LLMs. Furthermore, to harness the full potential of these LLM assistants for seeking help for their software tasks, we need more insights on where and how users struggle with these LLMs \cite{MCCRACKEN_2023}. Our study contributes new knowledge on how non-AI expert end-users employ LLMs' generated software guidance assistance in accomplishing tasks for feature-rich applications by comparing our own implemented, SoftAIBot, an LLM optimized for particular domain context (e.g., software documentation) with the Baseline ChatGPT. 

\subsection{Prompt-based interactions} 

To leverage the potential of LLMs, a lot of the focus in HCI and AI research is turning to \textit{prompt-based interactions} as users generally have to provide input or queries in the form of prompts that are then processed or responded to by a conversational AI system  \cite{Johnny}. Recent studies on the usability of prompt-based interactions \cite{Advait, Xu, Barke} reveal that prompts have a significant impact on pre-trained language models' ability to produce desired outputs, even though the prompts themselves are simple textual instructions for the task at hand \cite{Johnny}. For example, Advait et. al's (2022) \cite{Advait} study on the use of LLM-assisted tools for programming tasks revealed that the crucial concern is crafting effective prompts that elevate the probability of an LLM model to generate efficient code. Thus, the big challenge for end-users, especially novices and non-AI experts, is to define the appropriate prompts and learn prompting strategies to get the desired assistance from these LLMs. 

Unlike traditional help-seeking mediums that rely on keyword matching, prompt-based interactions within LLMs offer human-like language capabilities \cite{vera_liao}, which is unique, but can also be unreliable.  This unreliability comes from the biases (e.g., hallucinating and non-deterministic output) inherent within prompt-based interactions of LLMs. Considering LLMs are a tremendous leap from traditional help-seeking mediums that most users are familiar with, there have been calls to investigate users' mental models as they interact with LLMs \cite{vera_liao}. This becomes necessary when seeking assistance for feature-rich software tasks, where there is an interplay between the mental model of LLM vs the software application \cite{vera_liao, Kiani_help}. Recent studies have focused on understanding users' prompting strategies and proposing a catalogue of prompting guidelines \cite{prompt_catalogue, Suhridpalsule_2023, Shieh_2023} for allowing users to craft better prompts and seek desired LLM assistance. However, the use of these prompt guidelines in practice and their effectiveness remains unclear. Our study complements the existing research by observing users with prompt-based interactions and assessing the efficacy of prompt-based guidelines and integration of domain context in enhancing LLM assistance for software tasks.

\label{related_work1}

\section{Method: Controlled Experiment and Follow-up Interviews} \label{method}
We conducted a two-part user study with 16 users that consisted of a controlled experiment and follow-up interviews. Our main goal in this study was to investigate the effectiveness of two recent advancements: 1) prepending prompt guidelines to user prompts to enhance the accuracy of LLM output, as advocated by OpenAI, Microsoft and others to enhance Generative AI tools \cite {Suhridpalsule_2023, Shieh_2023}; and, 2)  directly integrating domain context (e.g., software documentation) into LLM assistants (e.g., as demonstrated in Copilot in Microsoft 365 applications \cite{copilot} and Firefly in Adobe applications \cite{Adobe_2023}, etc.) to enhance the relevance and accuracy of LLM output for software-related tasks. For our investigation, we implemented both of these advancements in a new GPT-4-based assistant, which we call \textit{SoftAIBot}, that offers in-context prompt guidelines (See Figure \ref{soft_ai_1}) and enhances the LLM output by making it specific to the feature-rich application (See Figure \ref{soft_ai_1}.c). For comparison, we also implemented \textit{Baseline ChatGPT} based on the pre-trained state-of-the-art LLM assistants, ChatGPT. (SoftAIBot is explained in more detail below.) \change{The prompt suggestions were not included in the Baseline ChatGPT because it was a control condition for the experiment.}

Based on our research questions, we derived the following hypotheses for users seeking software help:

\begin{itemize}
    \item \textbf{H1:} Users will perceive SoftAIBot as being more accurate than Baseline ChatGPT.
    \item \textbf{H2:} Users will perceive SoftAIBot as being more relevant than Baseline ChatGPT.
    \item \textbf{H3:} Users will trust SoftAIBot more than Baseline ChatGPT. 
    \item \textbf{H4:} Users will find the output provided by SoftAIBot to be easier to apply in the software application than Baseline ChatGPT. 
\end{itemize}

\subsection{Participants}
We recruited 16 participants (9F$|$7M) for our study, focusing on non-AI expert users who had little to no prior experience or knowledge of ML or NLP. Our participants came from different backgrounds (CS, Engineering, Business, Arts) and professions (administrative services, business analytics, information designers, client services, students, and researchers). Participants were familiar with LLM-based assistant, ChatGPT (10/16), and a range of traditional chatbots (12/16) such as Siri, and Google Assistant (12/14). They had used ChatGPT before for text generation, text summarization, and programming tasks, but none of them used it for software tasks used in the study before. About half of the participants (7/16) had frequently used PowerPoint and Excel applications and the remaining were occasional users.  Our participants covered a range of age groups: 18-24 (25\%), 25-34 (62\%), 35-44 (13\%) and had different levels of education (2 Diploma, 4 Bachelor’s,  5 Master’s, 5 PhD). We recruited participants mainly from our university’s mailing lists and found additional participants through snowball sampling.

\subsection{Design and Implementation of SoftAIBot and Baseline ChatGPT}

 \begin{figure*}[!t]%
    \centering
    {\includegraphics[width=1\linewidth]{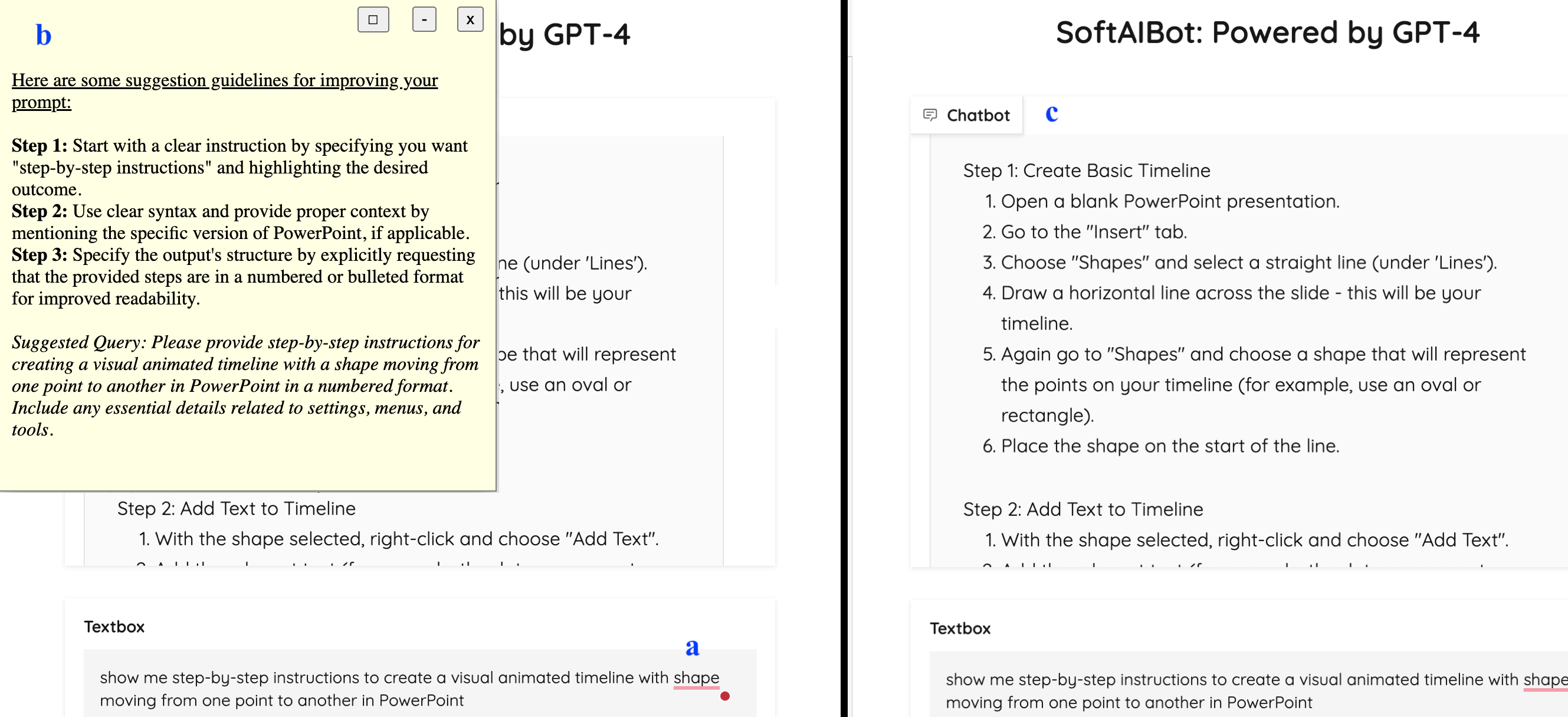}}%
    \caption{SoftAIBot integrates domain context via documentation and offers prompt guidelines to construct better prompts: (a) allows users to type in the prompt text and submit it; (b) generates prompt suggestions in-response to a user’s text (in this case, also shows a sample transformed query that users can directly use); (c) formats the response as step-by-step instructions optimized for particular software contexts, in this case PowerPoint. To see the contrast in LLM response, please see Baseline in Figure \ref{sample_task_ppt}.}%
    \label{soft_ai_1}%
\end{figure*} 

In this section, we describe the design and system implementation of our two interventions. 

\subsubsection{\textbf{SoftAIBot Intervention (GPT-4 with Prompt Guidelines and Software Documentation)} }

SoftAIBot LLM intervention suggests in-context prompt guidelines for constructing prompts while interacting with GPT-4 as well as generates guidance for particular software contexts.

\textbf{Automatic Prompt Guidelines:} The SoftAIBot UI interface (See Figure \ref{soft_ai_1}) lays out the various user interface components and receives the user’s prompt. Next, this user’s prompt gets transmitted to our custom API developed in Python connected to GPT-4 for generating prompt suggestions in-context to user’s prompt and a sample transformed query based on suggested guidelines that users can directly use. We used the prompt guidelines provided by OpenAI, Microsoft and others \cite{Suhridpalsule_2023, Shieh_2023}.  The generated prompt suggestions are displayed as an overlay (as shown in Figure \ref{soft_ai_1}.b) on the top left of the SoftAIBot interface via a \textit{Chrome} Extension which triggers with the click of “send” button component. These prompt suggestions appear after every prompt-based interaction initiated by the user.  To allow users freedom and control in accessing the prompt suggestions, we included the \textit{minimize}, \textit{maximize}, and \textit{close} options for either using the suggestion or simply closing it anytime during the interaction. 

\textbf{Implicit Domain Context Integration:}To optimize SoftAIBot for a particular software context (as shown in Figure \ref{soft_ai_1}.c), we leveraged GPT-4 augmented with corresponding software documentation. When a user submits the prompt, the relevant textual information or pertinent excerpts are extracted from software documentation by using Facebook AI Similarity Search (FAISS) index and vector search. Our custom API send this software context information along with user’s original prompt, together as a payload, to OpenAI’s GPT-4 8k to generate better tailored responses with reduced hallucinations. This process of using LLM with extracted relevant text is known as Retrieval Augmented Generation (RAG) Vector search \cite{Lewis_rag}. We used open source model BGE-smal-en \cite{muennighoff2022mteb} for producing chunks of the software documentation and transforming it to dense contextual vectors with 384 dimensions. \change{For the search approach used in RAG, we tried different approaches such as text chunking, and embedding models. The chunk size of 512 and BGE-smal-en approach provided us better context and fast retrieval.} Next, we searched this vector against the FAISS index to retrieve the most similar text vectors. Based on the indices of these similar vectors, we fetch the corresponding original text data from software documentation.\change{ We fed this extracted text from software documentation, along with the user’s query, as a prompt to GPT-4 (See Appendix \ref{technical} for details).}

\begin{figure*}[!t]%
    \centering
    {\includegraphics[width=1\linewidth]{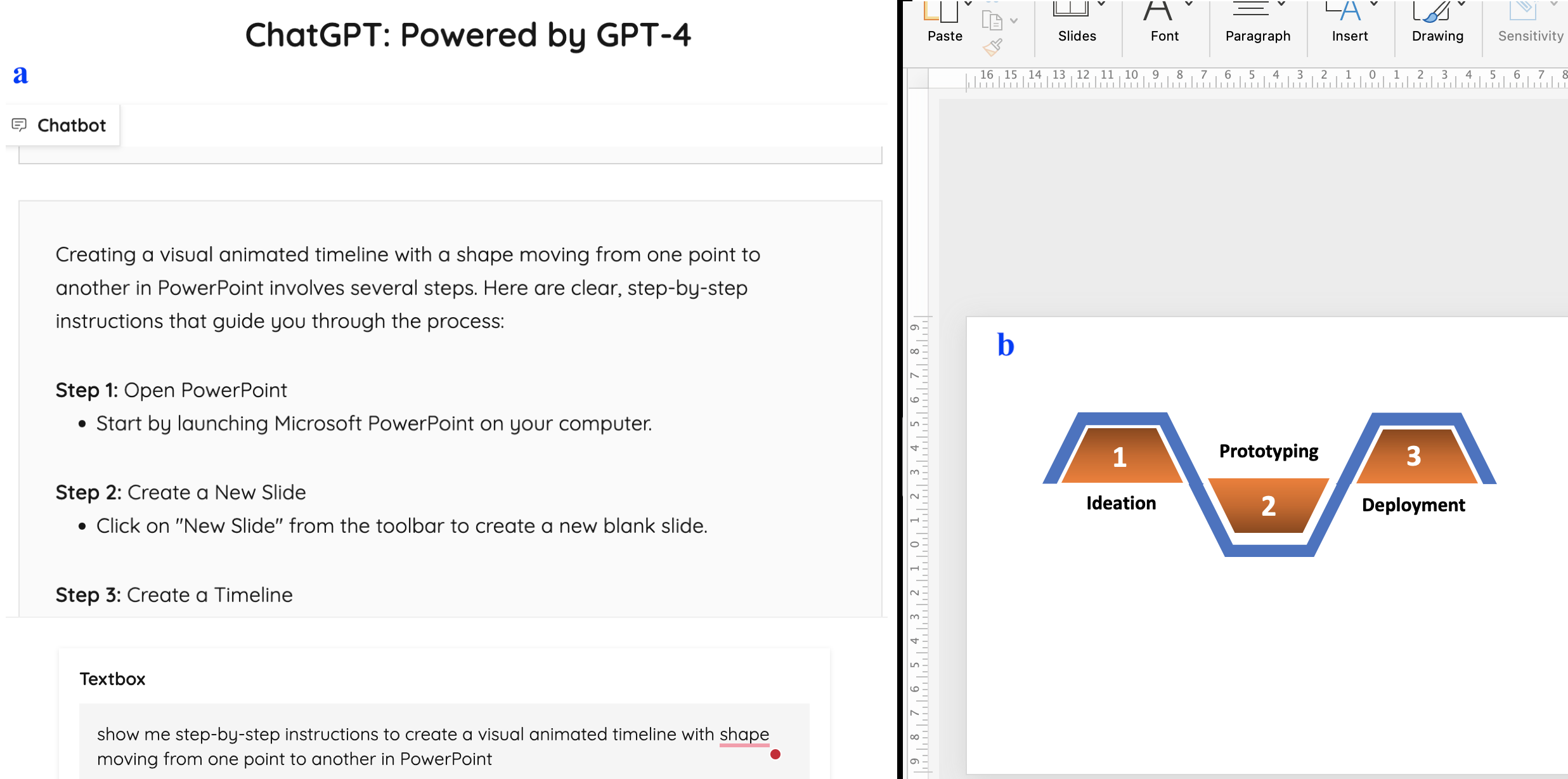}}%
    \caption{Overview of sample user task with PowerPoint application: (a) Users were asked to look up and use instructions from LLM intervention. In this case, \textit{Baseline ChatGPT} mimics the existing ChatGPT plus based on GPT-4, where users can type in their prompt in the textbox and LLM provide assistance to users for variety of tasks; (b) Use LLM assistance to develop shown project timeline in Microsoft PowerPoint that is visual and animated.}%
    \label{sample_task_ppt}%
\end{figure*}

\subsubsection{\textbf{Baseline ChatGPT intervention (ChatGPT plus)}}

 Our Baseline ChatGPT mimics the existing ChatGPT (See Figure \ref{sample_task_ppt}.a) plus based on GPT-4, where users can type in their query and LLM provide assistance to users for variety of tasks through out-of-the box multi-turn conversations using \change{OpenAI} GPT-4 8k \change{API} \cite{AI_2022, brown_llm}. To maintain the user's history and conversations when interacting with the OpenAI GPT-4 API, we \change{designed Python algorithm using chat completion objects \cite{Sanders_2023} (See Appendix \ref{technical} for details)}.
 
 To make the LLM UIs consistent for the experiment, we simulated both the interventions using Gradio framework \cite{Team_2023}, an open source framework, for developing intuitive web-based UI interfaces, making it easier to gather feedback, showcase results, and enable end-users to interact with LLMs during the user study.

\subsection{Choice of Application and Tasks}

To choose tasks and applications, we explored different productivity-related feature-rich applications (e.g., PowerPoint, Excel, Photoshop, Teams, etc.) popular among everyday novice users. After our initial exploration, we selected Microsoft PowerPoint and Excel to cover a range of tasks involving visual interactions, interactions involving application of statistical functions or formula and other visualization related tasks.

To assess the users' help-seeking approaches and observe any potential challenges when using LLM assistants for software help, we selected tasks that would require multiple steps for completion, and would necessitate multi-stage help and prompts (e.g., help within different steps needed for task completion). For example, one of the Excel tasks asked participants to use instructions from the LLM for analyzing and visualizing the predictive analytics of the sales values based on income using linear regression. Similarly, one of the PowerPoint tasks (See Figure \ref{sample_task_ppt}) asked users use instructions from the LLM to develop a project timeline in Microsoft PowerPoint that is visual and animated.


\subsection{Study Design and Procedure}

We used a within-subject design to minimize the effect of inter-participant variability. To eliminate order effects, we used a Latin Square counterbalancing \cite{raulin2019quasi} with 2 LLM conditions (total possible order= 4) to balance the order in which tasks were presented. During the experiment, each participant completed two tasks with each LLM interventions (4 tasks in total) . The participants performed all tasks using one feature-rich application, subsequently transitioning to a second feature-rich application, but the order of the tasks and associated LLM were randomized. 

Each study session began by introducing the participant to the LLM assistants, and provided some general tips to interact with the application (e.g., using the prompts). We conducted the study remotely through Zoom and participants were each given a \$15 Amazon gift card in appreciation of their time. Participants were provided instructions to install our LLM interventions via a Chrome extension. Next, participants completed a demographic questionnaire on their background and prior experiences with LLM assistants, chatbots and software applications. 

We presented each LLM intervention along with software application to the participant in a random order, one by one. We designed the complicated software tasks so that participants would be able to spend at least 8 minutes for completing each software task regardless of their familiarity and experience with the software application using given LLM intervention. After completing each of the 4 tasks, participants were asked to fill the post-task questionnaire hosted on the \textit{SurveyMonkey} to assess the overall experience seeking assistance from LLM interventions for software tasks along with their perceptions of accuracy, relevancy, ease of use, and trust of the assistance provided by LLM intervention. We encouraged participants to think aloud \cite{norman2013design} throughout the session and reminded participants that the study was seeking to understand how they seek assistance from LLM interventions for their software tasks rather than their performance or ability to master software applications or use the LLM.

Lastly, we conducted follow-up interviews to further probe any difficulties that impacted the use of prompt-based interaction and LLM assistance for software tasks, and any potential gaps in mental models about how LLMs work. Each session lasted approximately one hour and sessions were video and audio-recorded for transcription, and the participants were asked to share their screen through Zoom (only during the usability test).


\subsection{Data Collection and Analysis}
\label{data_analysis}
Throughout the study session, we recorded the participant’s screen and audio recorded their interview responses. We captured screen recordings to evaluate two key aspects: how users sought help from LLM assistants (e.g., formulated their prompts) and how they used the LLM output to complete the prescribed software tasks in Excel and Powerpoint. 

We used a combination of statistical tests and inductive analysis approach to make sense of the data captured from the user study. We ran Pearson’s Chi-square test for independence with nominal variable “LLM Interventions” (having two levels: SoftAIBot, and Baseline ChatGPT) and ordinal variable (having three collapsed levels: Agree, Neutral and Disagree) to quantitatively determine the significance of the results. 

\change{We used an \textit{expert-rating approach} where the experimenter (in consultation with all authors) analyzed the tasks performed by users in the software application and compared all of our metrics against the ground truth for all tasks. The ground truth in our context refers to the correct or optimal sequence of steps for task completion, the ideal application of software features, and the most relevant responses from the LLM to user queries. Our metrics included: }

\begin{enumerate}
    \item \change{\textit{Task Completion:} To measure the task completion, we evaluated how many of the software task steps (e.g., sequence of features/ functions) users completed using the LLM help.}
    \item \change{\textit{Task Accuracy:} To measure the participants’ success in applying LLM assistance to complete the study tasks accurately, we evaluated how accurately users identified the approach or functionality (e.g., macro, animation, motion paths, particular statistical function) from the LLM output and then applied the help using the corresponding software menu options and features in the software application. }
    \item \change{\textit{Accuracy and Relevance of LLM Assistance: }We compared queries and LLM response logs against pre-established ground truth to assess: a) how accurately the LLM provided instructions needed for successful software task completion; and, b) how relevant the LLM response was to the users’ input prompt.}
\end{enumerate}

Finally, to complement our experimental findings, we corroborated the data with participants’ think-aloud verbalizations and probed into the reasons behind users' decisions and identify any potential gaps in their mental models about how LLMs work. We used an inductive analysis approach \cite{corbin1990grounded} and affinity diagrams \cite{corbin1990grounded} along with discussions amongst the research team to categorize the interview findings and identify key recurring themes. \change{In particular, our coding approach for the inductive analysis considered reasons influencing users’ perception of how prompts impact output, recognition of accurate vs. hallucinated output, and difficulties in applying the LLM assistance to the software task.} 

\section{Results}

\subsection{Task Completion, Accuracy and Relevance of LLM Assistance}

\begin{figure*}[!t]%
    \centering
    {\includegraphics[width=1\linewidth]{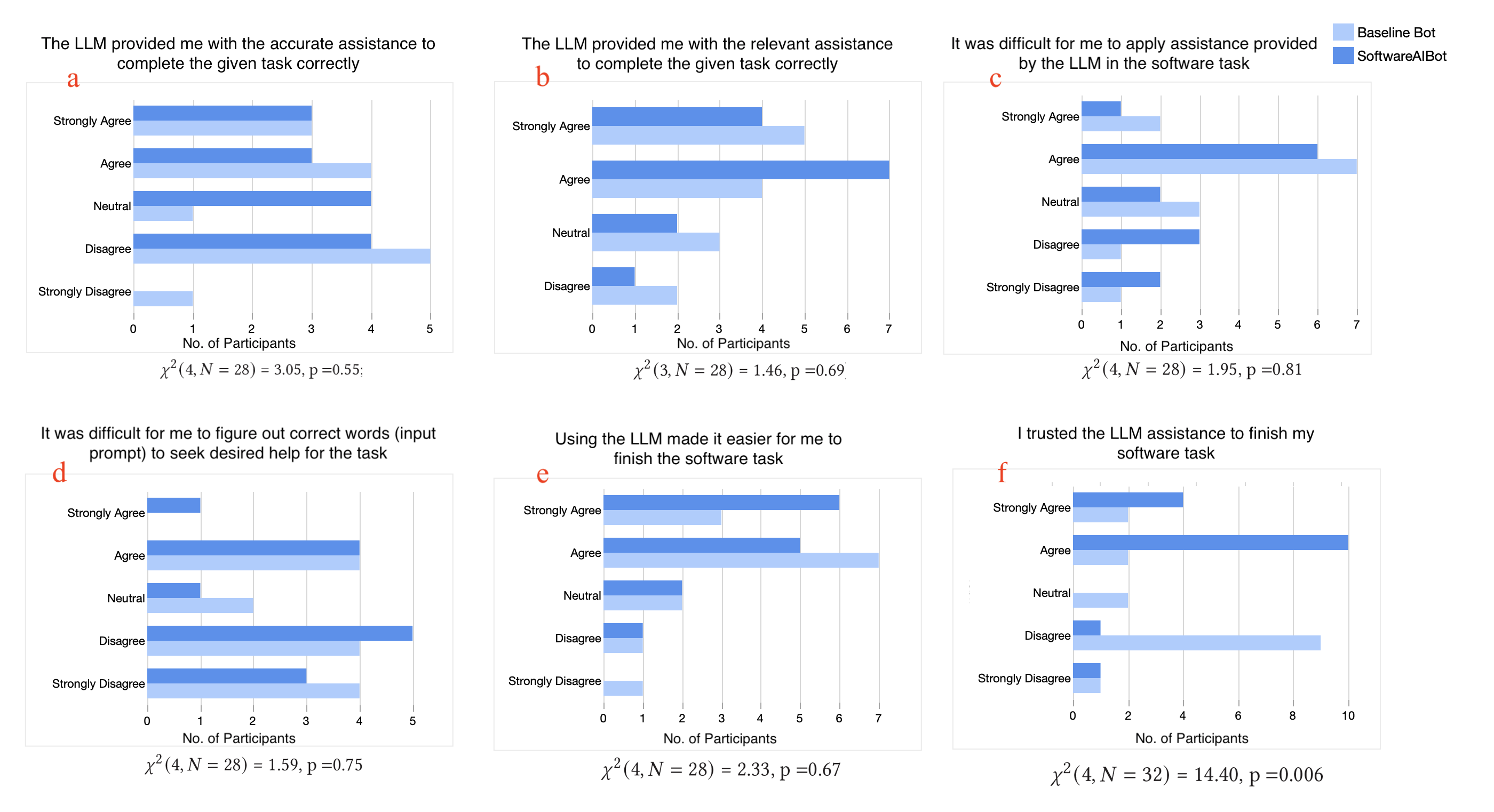} }%
      \vspace{-15pt}
    \caption{Overview of participants’ responses to post-task questionnaire. Pearson Chi-Squared test showed no significant difference for each metric across both LLM interventions for completing both Excel and PowerPoint tasks. Despite having low completion rate and low task accuracy, the majority of users perceived that they obtained accurate (a) and relevant (b) assistance from both LLM interventions. Still, the majority of participants (c) found it difficult to apply LLM assistance and instructions to the software application to complete their task; (d) Participants overall did not find it difficult to craft prompts; a few participants did indicate that they struggled to find the correct words for Powerpoint tasks that were more visual and interactive. Although expert ratings showed that users did not finish the task accurately, most users believed that it was easier for them to finish the task using both forms of LLM assistance; (f) The majority of users trusted both LLMs - this was surprising to see because expert ratings showed that both LLMs frequently provided inaccurate assistance. }  %
   
    \label{pearson_test}%
\end{figure*}

\textbf{Expert-rated Accuracy and Relevancy of LLM assistance: }
As predicted,  we found that the assistance provided by SoftAIBot was more accurate than Baseline ChatGPT for both PowerPoint (SoftAIBot: Mean=64.4\%; Baseline ChatGPT: Mean=37.5\%) and Excel (SoftAIBot: Mean=65.7\%; Baseline ChatGPT: Mean=45\%) tasks. These differences between accuracy scores and LLM interventions were significant for PowerPoint (\textit{t}(21.3) =4.0, \textit{p}=0.0006, two-tailed) and Excel tasks(\textit{t}(23.4) =3.7, \textit{p}=0.0011, two-tailed). Similarly, we found that the assistance provided by SoftAIBot had higher average relevancy score than Baseline ChatGPT for both PowerPoint (SoftAIBot: Mean=74.7\%; Baseline ChatGPT: Mean=44.4\%) and Excel (SoftAIBot: Mean=78.2\%; Baseline ChatGPT: Mean=55.4\%) tasks, with significant differences (PowerPoint: \textit{t}(24.5) =5.4, \textit{p}<0.0001, two-tailed; Excel: \textit{t}(24.8) =4.8, \textit{p}<0.0001, two-tailed).

\begin{figure*}[!t]%
    \centering
    {\includegraphics[width=1\linewidth]{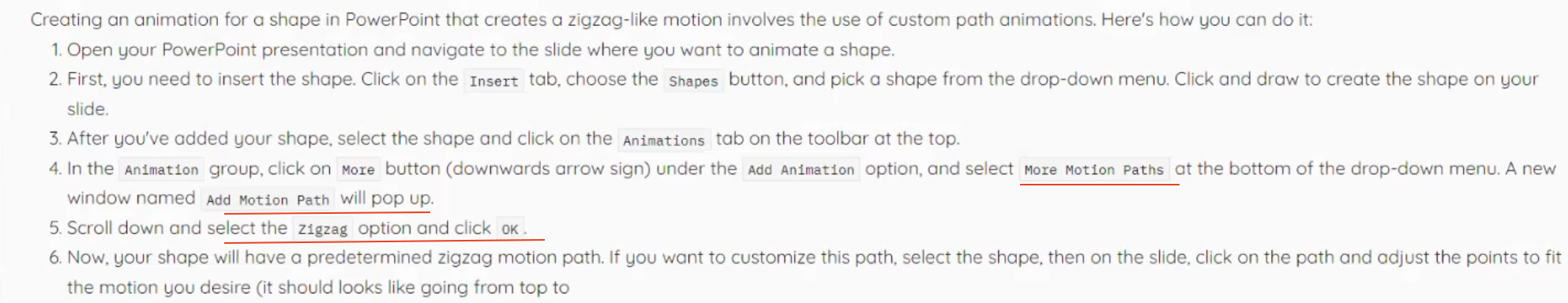} }%
      \vspace{-5pt}
    \caption{LLM Hallucination evidence (P15): In response to P15's prompt, “I want you to give me instructions on how to animate a shape that rotates from top to lower middle side and then come back up almost like a zigzag.”, Baseline ChatGPT generated the hallucinated response of Zigzag menu option (highlighted in red) which did not even exist in the software application.}  %
    \label{hallucination}%
\end{figure*}

Since the Baseline lacked relevant domain context, it typically failed to offer relevant and accurate steps available within the software (e.g., macros, animations, motion paths, etc.). Instead, it often provided references to features and functionality that did not exist. This phenomenon of the LLM providing information that is somewhat relevant to the user's query but not accurate to the users' intent for performing the task has been termed as an \textit{hallucination} \cite{vera_liao, bang2023multitask, borji2023categorical} in the literature (See Figure \ref{hallucination}). On the other hand, while SoftAIBot provided step-by-step instructions on how to implement the required functionality in the software, it also demonstrated instances of hallucination. Furthermore, it did not always provide specific and relevant instructions (for example, where to locate menu functions within the UI). 

In subsequent sections, we shed light on users' performance and qualitative perceptions of both LLMs, highlighting various inconsistencies and misconceptions among users that impacted their use of prompt-based interactions.

 \textbf{Task Completion and Task Accuracy: } None of the participants were able to completely finish either of the tasks in Powerpoint or Excel, even though each participant made full use of both LLM assistants (\textit{Baseline ChatGPT} and \textit{SoftAIbot}). On average, participants completed 35\% of the PowerPoint tasks (maximum= 50\% and minimum= 0\%) with Baseline ChatGPT and 45\% of the tasks (maximum= 60\% and minimum= 0\%) with SoftAIBot with no significant difference across the two LLM interventions (t (28.1)= 1.1, p= 0.28, two-tailed). 
 Similarly, participants completed 40\% of the Excel tasks (maximum= 55\% and minimum= 0\%) with Baseline ChatGPT and 55\% of the tasks (maximum= 75\% and minimum= 0\%) with SoftAIBot with no significant difference across the two interventions (t(28.7)=1.9, p=0.07, two-tailed).

Among the portion of the task completed by each participant, the accuracy scores were low. For the PowerPoint tasks, the average task accuracy score across all participants was 31.3\% using SoftAIBot (maximum= 50\%, minimum =0\%)
and only 17.2\% using Baseline ChatGPT (maximum= 50\% and minimum= 0\%). Although users achieved better task accuracy scores with SoftAIBot in comparison to Baseline, paired-sample t-test showed no significant difference across both interventions (\textit{t}(26.6) =1.7, \textit{p}=0.10, two-tailed), and we did not observe any order effects. The trend of non-significant accuracy persisted in Excel tasks as well (\textit{t}(29.9) =1.9, \textit{p}=0.06).

\textbf{Users' Perceptions of Accuracy and Relevancy of LLM assistance: } Even though task completion scores were poor across both LLM interventions, in the self-report data, surprisingly, the majority of users perceived assistance from both LLMs to be accurate (SoftAIBot: 12/16 participants; Baseline ChatGPT: 8/16 participants) and relevant (SoftAIBot: 13/16 participants; Baseline ChatGPT: 9/16 participants) in completing the PowerPoint tasks. Pearson Chi-Squared test showed no significant difference in \change{perceived} accuracy ($\chi^ 2 (4, N=28)$ = 3.05, p $=$0.55) and relevance of LLM assistance ($\chi^ 2 (3, N=28)$ = 1.46, p $=$0.69). Similarly for Excel tasks, there was no significant difference in \change{perceived} accuracy ($\chi^ 2 (3, N=28)$ = 3.96, p $=$0.27) and relevancy ($\chi^ 2 (2, N=28)$ = 0.29, p $=$0.86) across both LLM interventions.

In terms of other self-report data, such as users' perceptions of difficulty in applying assistance across both interventions, difficultly in figuring out correct input prompt, and ease of completing the task using LLM assistance, we did not observe any statistical difference across SoftAIbot and Baseline (See Figure \ref{pearson_test} for the detail statistical results and test on remaining metrics). The one exception was the perception of trust as, interestingly, we observed that most participants (14/16) trusted SoftAIbot more than the Baseline ChatGPT (4/16) for the PowerPoint tasks and this result was significant ($\chi^ 2 (4, N=32)$ = 14.40, p $=$0.006), but for the Excel task, there was no significant difference in users' perceptions of trust.

The key takeaways from our experiment were that while the expert rating showed that SoftAIBot performed better than the Baseline ChatGPT in producing more accurate results, users could not recognize the differences in accuracy and relevance among both LLMs. Furthermore, having more accurate and relevant LLM output did not impact task completion nor task accuracy across both software applications. Next, we use our qualitative findings to explain factors that impacted user performance and highlight  some of the key challenges that users experienced in figuring out appropriate prompts and applying LLM assistance for different software tasks. 

\subsection{Inconsistency in Prompt-based Interaction}
\begin{figure*}[!t]%
    \centering
    {\includegraphics[width=1\linewidth]{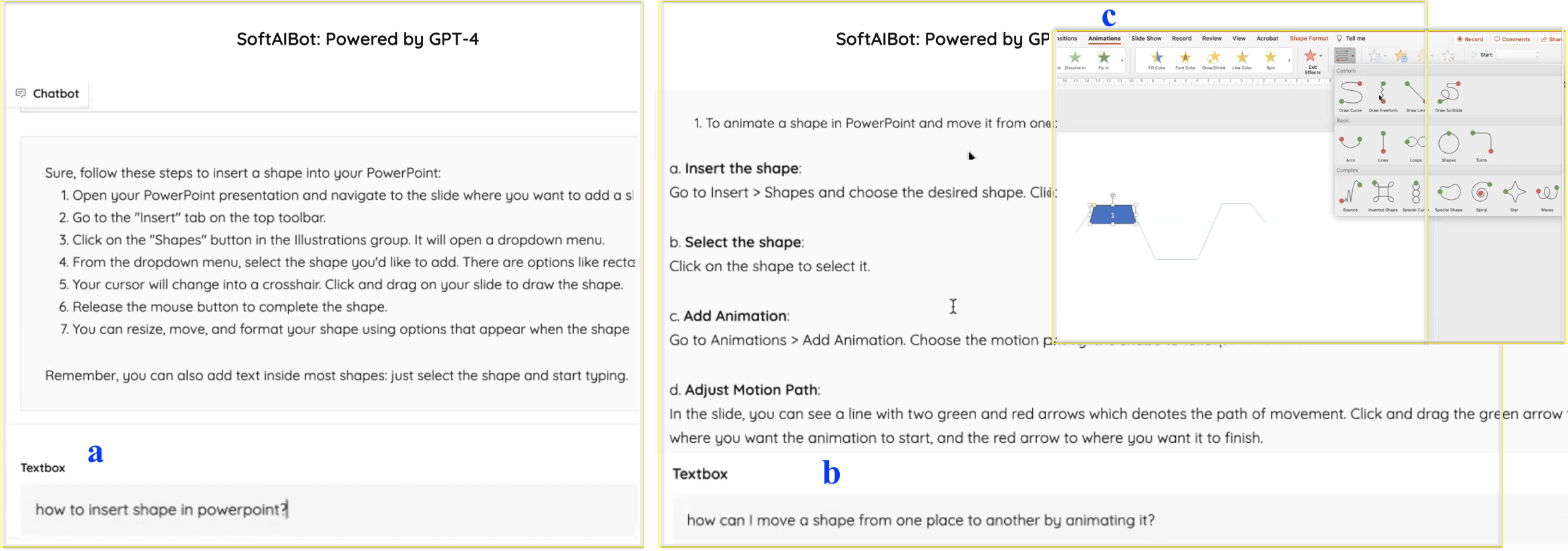} }%
      \vspace{-5pt}
    \caption{Example of successful prompting by Participant P02. They started by breaking down the task, beginning with asking steps to (a) implement shape, followed by (b) animations in PowerPoint task. User prompted SoftAIBot as a step-by-step process asking steps for each functionality at one time (P02). }  %
    \label{succesful_prompting}%
\end{figure*}

 Prompt guidelines and prompt engineers usually suggest that breaking down the task and prompting it as a process for LLM can usually lead to desired LLM outputs. However, with the exception of 2 participants in our study who followed this approach and were successful (See example of Participant P02 in Figure \ref{succesful_prompting}), all of the other participants were inconsistent and varied in how they constructed prompts, failing to leverage the in-context prompt guidelines. 

\begin{figure*}[!t]%
    \centering
    {\includegraphics[width=1\linewidth]{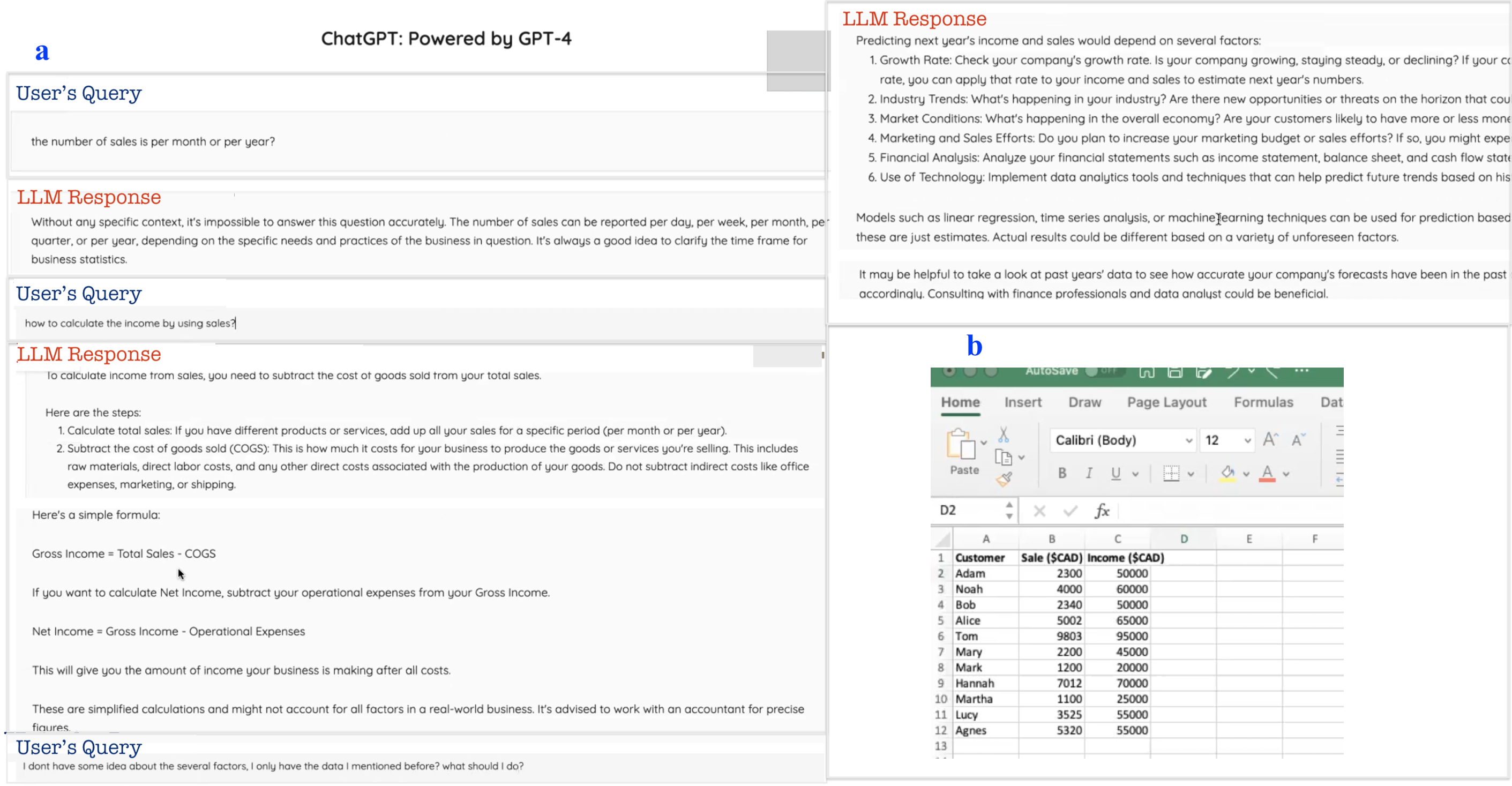} }%
      \vspace{-5pt}
    \caption{Unsuccessful Prompting by using keyword-based approach with Baseline ChatGPT: (a) P07 prompted the LLM using the keywords interpreted from the task, and struggled in getting an relevant response and went through several rounds of clarifications with the LLM;  (b) Eventually, user could not even get started and failed to perform the task on the software application (P07). }  %
    \label{unsuccessful_prompting}%
\end{figure*}

\textbf{Constructing Prompts as Search Queries:} \label{constructing_prompts} Most participants (10/16) started with a generic \textit{“how to do the [task]}” prompt, such as (e.g., “how to animate in Microsoft PowerPoint” (P14), “\textit{how to do correlation in Microsoft excel}” (P04)). Most users were translating their mental model from other query-based systems, relying on similar queries they would issue on Google:\textit{ “I think it [LLM] relies on the keywords that I am giving...at the beginning I just formulated a very vague question because it's easier to get started with the vague question and then I can refine it as I go.”} (P08) 

Even after users experimented with different phrasings, they could not understand why the LLMs were producing nearly identical responses. These participants did not have an accurate mental model of how LLMs work and did not appear to recognize the impact of the prompt text on the quality of the LLM output: 
“\textit{I think it works same as a search engine. It has a back end and it takes your question through tons of data...it tries to give you an answer with all of that data that it has in the back end..it's so quick that it goes through it within nanoseconds..”} (P15) 

Some participants (4/16), without even interpreting the task, just copy-pasted the entire task instructions along with some data sample (e.g., for Excel tasks) in hopes that the LLM would simplify the task and provide some instructions. However, both of these approaches were not that successful (as shown in Figure \ref{unsuccessful_prompting}), as users had to ask follow-up questions on figuring out the correct steps. For example, when P07 (Figure \ref{unsuccessful_prompting}) prompted the LLM using the keywords interpreted from the task, they struggled in getting an relevant response and went through several rounds of clarifications with the LLM. Only 2/16 participants who broke down the task and drafted prompts as a process for SoftAIBot LLM obtained desired LLM outputs (See example of P02 in Figure \ref{succesful_prompting})

\textbf{Using Trial and Error Due to the Vocabulary Problem:}  The majority of participants (11/16) also struggled in crafting prompts because they did not know how to express their task intent using software-specific terminology (often termed as the vocabulary problem \cite{Furnas}). This was especially prevalent during the PowerPoint tasks that involved references to different visual and interactive elements and participants frequently engaged in long trial-and-error episodes. Participants found it challenging to articulate their intended actions accurately and frequently blamed themselves: \textit{“I did not know how to describe those visual graphics in PowerPoint...I said words like flip and move but I am not sure if it was right for these kind of tasks . It took me a lot of time to try to understand which menu to select and not being sure if my prompt was ok or not… If the problem was my prompt or my system or the problem was the solution provided, I was not sure which one of them was making mistake.”} (P05)

\textbf{Ignoring Prompt Guidelines: } Contrary to our hypothesis, only a handful of participants (5/16) employed the in-context prompt guidelines provided by SoftAIBot within the context of their queries and the majority simply ignored them. Participants expressed that the prompt guidelines were “not necessary” and “not useful” as it was faster for them to iterate on their own prompts. This behaviour was similar to the phenomenon commonly referred to as the “active user paradox” \cite{Carroll}. In fact, about half of the participants (7/16) were confident that they already possess the knowledge and experience required for generating prompts: \textit{“I would not say it is hurtful to have it [prompt guidelines] but is not necessary. I knew what to search for. I do not think prompt guidelines would have helped…”} (P06)  Some participants (4/16) noted that they were confident about crafting their own prompts because the LLMs were able to accept “any input” and generate corresponding output.   If needed, they can revisit the generated response and assess their prompt alignment with their intended outcome for further improvement: \textit{“It is not difficult to figure out words because it [LLM] was accepting anything I typed. I can go back and verify whether that is what I need.”} (P11)

\subsection{User Perception of LLM Assistance}

Some of the surprising results from our experiment were that users did not recognize the differences in accuracy and relevance between the two LLMs nor were they able to leverage the assistance to complete the software tasks accurately. Below, we discuss some factors that shaped users' perceptions and use of both LLMs. 

\textbf{LLMs' Ability to Generate Response Created an Impression of Credibility: }  Most participants (13/16) perceived that both LLMs produced relevant responses because in contrast to systems like Siri and Google, which do not always provide response for every user's query, both LLM consistently generated a reasonable response matching the user's prompt. Thus, most participants formed an impression that the LLM outputs were credible: \textit{“... it [LLM] gave me what was relevant to my query...like some steps to find those options. This bot [LLM] was more detailed even than Siri. When you ask Siri, it's not giving what exactly I am looking for and keeps on giving me some different options...which is unnecessary about the topic. But in this one [LLM], what you put that's what you're getting. So the output is like, 90 to 95\% near to what you just asked...so that made me trust it.”} (P12)

\textbf{Difficulty in Applying LLM Assistance to the Software Application: } In addition to the tensions in forming an accurate mental model of both LLMs (as described in Section \ref{constructing_prompts}), our participants who were infrequent users of Powerpoint and Excel also struggled because they lacked an accurate mental model of these applications. We observed that participants were quick to blame themselves for not being able to apply the instructions given by LLM due to lack of their familiarity with software: \textit{“For some reason, I got what it [LLM] tells me but...maybe it’s giving me correct step, but I am not able to apply to Excel because I’m not a regular user. I need to figure out [myself] how to use this formula; where can I put my formula and how to apply it.”} In comparison to SoftAIBot, users felt that the Baseline ChatGPT generated more generic or vague responses which users struggled to apply to software tasks. In fact, 9/16 participants began to doubt the credibility of Baseline ChatGPT: \textit{“Unless you apply the steps, you do not know whether that [Baseline ChatGPT] works or not.”} (P06)

In cases where the LLM output was actually relevant and accurate, our participants still struggled to locate menu options and apply LLM instructions. They explained that it was due to the lack of visual cues or guidance that are typically provided in tutorial videos and other forms of visual help: \textit{“...the instructions were here, but sometimes it gets difficult to use...If I get an image or graphical help along with screenshots of where to look for particular option, I could have made it slightly faster.” }(P11) Due to this difficulty in locating and applying LLM assistance, few participants (6/16) came up with different theories on whether LLM output did not considered their system version, as commented by P04:  \textit{“It [LLM] did not provide instructions for the version of the software I'm using. Sometimes it's little bit difficult to apply the assistance, sometimes I would not find the button he [LLM] asked me to look for.” }(P04) 
 
When the LLMs hallucinated or produced an incorrect set of instructions, we observed that more than half of the participants (10/16) could not map the (wrongly generated) LLM output to the intended features and menu functions in the application. As a result, they formed an incorrect mental model of the software application's user interface. 
Instead of having awareness of this LLM's bias of hallucination, these users felt burdened to make LLM instructions work:  \textit{“For someone who is very new to PowerPoint...I could not find like the stuff, the menu names [ZigZag] that it was mentioning. It was not there in my application. I am not sure why all the burden of [finding] came to me.”}(P15) Only a few participants (3/16) were able to recognize that the LLM does not always give factual information. They compared LLMs with Stack Overflow and Google search which they considered to be more credible than LLMs: \textit{“ ...if it's a Stack Overflow, I would know that it's just one person's comment so I have ways to verify how trustable that instruction is...with Google, I have that much control over the source of information. But with LLM, I have no way to verify that at first sight. I have to follow it and, and decide on my own if it works or not. I prefer to be able to verify the credibility of a solution before actually going through the steps and putting more time to it.”} (P07)

\subsection{Coherent LLM Output Leads to Blind Faith}

\textbf{The presentation of LLM output fosters trust:} The most surprising finding from our study was that after receiving an output from the LLM, most users blindly followed the provided steps without a critical evaluation of the output's veracity. For example, as illustrated in Figure \ref{hallucination}, even when Baseline ChatGPT produced the hallucinated response of the \textit{Zigzag menu} option which did not even exist in the software application, P15 still demonstrated unwavering trust of this LLM. The boundary between right and wrong for the participants while seeking LLM assistance, even in cases where it might yield incorrect results, was blurred as both LLMs consistently produced output that is coherent and in plain English, which enhances its perceived credibility. Especially for SoftAIBot, most participants (14/16) assumed that LLM is correct just because it generated a well-formatted response in response to their query: \textit{“I trust the system because once I get the match answer from him [LLM]...makes me feel he [LLM] is helping and he is better than me.”} (P03)  Another user who followed SoftAIBot's step-by-step instructions for PowerPoint's visual tasks commented: \textit{“I was able to trust it [SoftAIBot]  because I liked the way it gave these steps. SoftAIBot is more specific and gave me step-by-step sort of directions...it was user friendly and easy to follow.  I could find all the steps that it [SoftAIBot] was referring to. Because this is what you need when you ask AI for help. You look for baby steps.” (P14)}

\section{Discussion}
\change{We have contributed insights into how novices employ new-generation LLM assistants to seek software help, highlighting many of the challenges that users face while crafting prompts, comprehending how prompts bias LLM output, and mapping the LLM-suggested steps to software.} Our key findings suggest that even though SoftAIBot outperformed the Baseline ChatGPT in providing relevant and accurate software-related assistance, there were no significant differences in users' task completion rates or task accuracy scores between the two conditions. Notably, \change{as opposed to our hypotheses H1-H2,} there was no difference in users' perceptions of accuracy and relevance for both LLMs, and users mostly failed to recognize instances where the models provided incorrect answers, including hallucinations. Our qualitative findings \change{further shed light on our research questions and }reveal a lack of awareness among participants regarding LLMs' biases and limitations. Participants attributed their inability to complete a task and locate suggested features in the application to their personal shortcomings rather than recognizing instances of LLM hallucinations offering nonexistent options. Additionally, users misunderstood the prompt text's influence on LLM output, likening prompts to traditional search engine keywords. User perception of LLM responses as contextually relevant stemmed from a bias towards the query context. Importantly, our study highlights the risks of undue trust in LLMs, as users frequently exhibited unwarranted confidence in LLM-generated responses due to their human-like nature and consistent, contextual relevance, distinguishing them from traditional chatbots or virtual assistants like Siri.

The implications of our research extend beyond the immediate findings and have far-reaching significance for the broader IUI research community. Our observations highlight the need for end-users to exercise caution and critical thinking when relying on LLMs for software-related assistance. In an era where LLMs are increasingly integrated into various facets of daily life, from virtual assistants to content generation tools, understanding user perceptions and misconceptions about these systems is imperative. By shedding light on the lack of awareness regarding LLM biases and hallucinations, our study calls for a fundamental reevaluation of the way we design, deploy, and educate users about AI-powered assistants.

We now reflect on our key insights and highlight opportunities for designing LLM assistance for feature-rich software tasks while promoting transparent, responsible LLM interfaces to enhance user understanding and mental model formation. Our findings will be valuable for IUI and HCI researchers, interface designers, developers and others working on LLM-powered assistants.

\subsection{Integrating LLM help into feature-rich applications}

Our results demonstrate the value that LLM-based assistants, such as ChatGPT, can provide in generating relevant software-related assistance within a single platform. Unlike traditional help-seeking resources and chatbots (e.g., Google search, blogs, Siri, etc.) where users have to assimilate help content through multiple outlets, our participants appreciated receiving relevant detailed instructions by typing in a prompt. Having said that, our findings resonates with the speculations of other researchers \cite{Kelly_2023} that Baseline GPT-4 is not meant to provide assistance for all types of tasks. Our SoftAIBot, that was optimized for particular feature-rich software guidance context, performed better and generated more relevant and accurate step-by-step software assistance. In our approach, we employed Retrieval Augmented Generation (RAG) on standard software documentation. Future developments could involve \textit{instruction tuning} \cite{zhang2023instruction}, which includes pairing more specific instructions with the software-specific steps and correlating this with expected output. The onus should transition from users to software developers and customer support to create such instructional pairs for ensuring them to be crafted in a manner that allows general-purpose LLMs to be fine-tuned \cite{Liu_finetuning} for generating user-centered and optimized software guidance.

While SoftAIBot generated relevant and accurate software assistance compared to Baseline ChatGPT, there were obvious limitations as users were not able to leverage this information to complete the software tasks accurately. In particular, users found the textual LLM output to be limiting compared to other visual-based help-seeking mediums. For example, software instructions on YouTube allow users to follow procedural steps “as is” without needing to verify and locate specific features. However, with LLMs, users had difficulty in mapping the LLM output to features in the software, especially in cases where the LLM was hallucinating and referred to non-existent features. Video-based help-seeking mediums should not be dismissed as instructional tools. Instead, to mitigate the issue of locating the exact instructions that users experience with videos, there is an opportunity for technologies like ChatGPT to aid in video summarization tasks and to extract more relevant snippets from videos \cite{Fraser_multimodal}, enhancing the ease of locating specific information. 

\subsection{Transparent and Responsible Interface Design of LLMs}
At a more fundamental level, our study raises some caution: while developers and researchers are investing in improving AI models and how LLMs can provide context-specific guidance, users may not always perceive these enhancements as substantial improvements in accuracy and relevance. \change{Recent literature has already raised concerns about users' over-reliance on AI systems, such as in the context of AI-based maze-solving tasks \cite{Vasconcelos}. Although the landscape of user behaviors and mental models is more multifaceted with LLMs, our study demonstrates a similar phenomena of over-trust with LLMs.  Furthermore, we extend prior works by revealing the } nuances in users' mental models of the LLMs \change{triggered by the inherent biases of LLMs}, leading to overtrust and users' failure to recognize erroneous or hallucinated output: “\textit{Because it is AI, how can it be wrong?  I am going to stop using my brain as I literally gave it gibberish and still it works. I [will] doubt myself before doubting AI}.” (P15)  Such overtrust in LLM assistants can be dangerous, especially for novice users who do not have familiarity with the underlying powerful AI technology. \change{These findings} from our study underscore the complexity of user interactions with AI and highlight the need for more transparency in addressing users' expectations and misconceptions. This can be as critical as advancing the underlying AI technologies and it is essential to design interfaces that are more transparent and responsible \cite{Sun_explainable}. There is need to consider more innovative user-centered solutions for mitigating bias and enhancing transparency in AI systems, thereby contributing to the responsible and ethical development of AI technologies.

To enhance the transparency of LLMs among end-users, one approach could be to embed interpretability within these systems. By articulating \textit{why} and \textit{how} LLMs derive specific recommendations, users can gain perspectives into the underlying mechanisms and trust the provided instructions with a higher degree of certainty. One possible direction is through the illustration of confidence percentage scores \change{for each LLM instruction as these scores have shown to enhance the perception of transparency and trust \cite{Khurana}. Other potential direction is through explainability techniques \cite{passi2022overreliance} (e.g., visual example-based explanations \cite{Carrie, Khurana}) that can indicate why the system did what it did and verify an AI’s recommendation \cite{fok2023search} by demonstrating} the similarities between users' intent and examples in the training set \cite{Khurana, Toby_multimodal, Toby_2, Carrie}. Training datasets of these LLMs needs to be designed such that confidence scores \change{or visual examples} are part of the dataset to enhance transparency within the LLM technologies. Such innovative advancements in LLM development not only pave the way for enhanced user interaction but also ensure that the model's suggestions are verifiable and reliable. 


\vspace{-8pt}
\subsection{Bridging the Gap Between Mental Models and LLM Interfaces}

With the rapid pace of innovations in Generative AI and LLMs, there is need for HCI research to focus on understanding user perceptions and users' mental models of LLM-based assistants for software help-seeking. Our study \change{complements existing works highlighting usability issues in crafting prompts \cite{Advait, Xu, Barke}} and provides initial insights into the gaps in end-users' mental model when using prompt-based interactions \change{in context of software help-seeking}. The affordances of LLMs can be misleading as they are designed to be walk-up-and-use and support natural language interaction. However, similar to insights from recent work on non-ML expert designers prototyping ML apps \cite{Johnny}, we also found that crafting effective prompts is cumbersome, especially for non-AI experts. Our study contributes new knowledge: non-AI expert end-users of ML/LLM applications must adapt their existing mental models from traditional help-seeking mediums to understand the new interface of LLMs. To bridge the gap between users' mental models and LLM user interfaces, there is an urgent need to leverage strategies such as think-aloud studies to further understand nuances in users' mental models of LLMs \cite{norman2013design}. \change{Although we did not see significant individual differences in our sample, it may be worth investigating how different sub-groups of software users might benefit from LLM help seeking.} Our study demonstrates that we cannot assume users will intuitively grasp the capabilities and limitations of LLMs. There is a clear need for comprehensive user training and education and clear communication about how Generative AI systems operate.

With software help-seeking, the challenge for end-users lies not only in flawed mental models of LLMs but also in the absence of a clear understanding of the underlying software application. \change{Our research complements the emerging work in this space, being novel in documenting the challenges users encounter with LLMs for software help tasks, including their mental models and overtrust, which impacts both their utilization of LLMs and perception of software features, regardless of LLM optimization or explicit prompt guidelines (SoftAIBot).} Users found it difficult to understand, map, and apply LLM instructions to software features. To address this, enhancing the UX design of LLM interfaces by highlighting relevant software UI sections during onboarding can improve user interaction, particularly for feature-rich software\cite{Khurana, Toby_2, Toby_multimodal}. \change{Exploring the interplay between users' understanding of LLMs and the underlying software presents new human-AI design possibilities.} Overall, our findings show that LLMs optimized for generating software specific guidance (e.g., Copilot \cite{copilot}) and embedded inside feature-rich applications, could be promising for learning and using complex features. Once these systems become available, future studies can compare our findings from SoftAIBot with such systems and further investigate the level of guidance and automation that may be appropriate for LLMs generating software guidance.

\section{Limitations}

In this paper, we experimented with two LLM-based assistants to explore how end-users make use of them for software tasks related to feature-rich applications. While our findings shed new light on users struggles in employing LLMs for software help-seeking, some caution should be used in interpreting our results. For instance, our findings could be constrained by the specific applications utilized during the experimentation. Whether our findings would generalize beyond the state-of-the-art LLM implementations used in the study should be investigated in future work. Given the rapid evolution and variability among modern LLMs, the outcomes may be constrained to currently available LLMs. Although we recruited end-users who were non-AI experts, we did not control for other individual differences, such as expertise or familiarity with underlying feature-rich software applications. Future studies should conduct experiments and more qualitative studies with larger and varying populations to better capture these individual differences.

\section{Conclusions}

In this research, we investigated the effectiveness of the LLM-generated software guidance and prompt guidelines, by comparing the Baseline LLM assistant with our own implemented SoftAIBot in providing accurate and relevant assistance for software tasks. Our results highlight the challenges users faced in following LLM assistance and point to instances of users attributing flaws to themselves instead of recognizing LLM biases. Our study highlights the pressing need for interdisciplinary collaboration among researchers, designers, developers, and educators to bridge the gap between user expectations and AI realities. By addressing these challenges head-on, we can foster a future where AI systems are not only more powerful but also more comprehensible and accountable to their users, ultimately facilitating human-AI interaction across a range of domains.

\begin{acks}
We thank the Natural Sciences and Engineering Research Council of Canada (NSERC) for funding this research.
\end{acks}

\bibliographystyle{ACM-Reference-Format}
\bibliography{99_refs}
\newpage

\end{document}